\documentclass[aps,twocolumn,superscriptaddress,showpacs,amsmath,amssymb]{revtex4-1}

\usepackage{graphicx}
\usepackage{dcolumn}
\usepackage{bm}
\graphicspath{{pic/}{pict/}{}}
\newcommand\leqt[1]{\protect\label{#1}}

\newcommand\lsect[1]{\protect\label{sect:#1}}
\newcommand\rsect[1]{\ref{sect:#1}}

\usepackage{hyperref}
\hypersetup{pdfstartview={FitH},pdfpagemode=None,
            colorlinks,linkcolor=blue, citecolor=blue,
            bookmarks=true, bookmarksopen=true, pdfnewwindow=true}

\begin{document}

\title{Slow phonon vortices and defect modes in periodic nano-waveguides}

\author{Yue Sun}
\email{yue.s@anu.edu.au}
\affiliation{Nonlinear Physics Centre, Research School of Physics and Engineering, The Australian National University, Canberra, ACT 2601, Australia}
\affiliation{Laser Physics Centre, Research School of Physics and Engineering, The Australian National University, Canberra, ACT 2601, Australia}

\author{Anton S. Desyatnikov}
\affiliation{Nonlinear Physics Centre, Research School of Physics and Engineering, The Australian National University, Canberra, ACT 2601, Australia}
\affiliation{Department of Physics, School of Science and Technology, Nazarbayev
University, 53 Kabanbay Batyr Ave., Astana 010000, Kazakhstan}

\author{Andrey A. Sukhorukov}
\affiliation{Nonlinear Physics Centre, Research School of Physics and Engineering, The Australian National University, Canberra, ACT 2601, Australia}

\date{\today}

\begin{abstract}
We identify a broad class of phonon modes with persistent vortex fluxes at arbitrarily slow propagating velocities in periodic nano-waveguides. Such phonon vortices are associated with the split band-edges in dispersion dependencies, which can be engineered by waveguide design. Modulations introduced in such waveguides can support a pair of defect cavity phonon modes with an arbitrarily small frequency splitting. These features can find applications for sensing and nano-manipulation.
\end{abstract}

\maketitle



%

\section{Introduction}

Periodically modulated nano-waveguides offer unique potential for controlling the wave propagation velocity and trapping cavity modes at specially introduced defects, based on dispersion modification and formation of band-gaps through Bragg reflection effect. Remarkably, it is possible to design a structure to simultaneously control waves of different physical origins, including photons and phonons~\cite{Pennec:2011-41901:AIPA, Gomis-Bresco:2014-4452:NCOM, Marchal:2012-224302:PRB,Graczykowski:2015-075414:PRB}. Such phoxonic crystals can facilitate strong acousto-optic interactions, opening new opportunities for frequency conversion, sensing, nano-manipulation, and other applications~\cite{Aspelmeyer:2014:CavityOptomechanics}.

The wave interactions can be enhanced for frequencies close to band edges. In this regime the group velocity is strongly reduced, which can lead to the increase of wave energy density~\cite{Khurgin:2009:SlowLight}. Importantly, the actual dynamics of interactions depends on the nano-scale features of the mode profile. In particular, the structure of energy fluxes can play a critical role in context of nano-manipulation.

In this paper, we demonstrate an approach for obtaining persistent vortex fluxes for phonons, in the regime of slow group velocity propagation. We show that this is realized when the phonon dispersion is specially engineered to feature so-called split band edge (SBE), which is associated with wavevectors inside a Brillouin zone. The phenomenon of vortex flows is conceptually analogous to the corresponding features of optical modes in the SBE slow-light regime~\cite{Sukhorukov:2009-94016:JOA}. We establish that structure of vortices for slow phonon modes involving both transverse and longitudinal oscillations can be explained by establishing correspondence with the theory of optical polarization singularities~\cite{Dennis:2009-293:PROP}.

Furthermore, we demonstrate new possibilities for engineering the acoustic defect cavity modes, by employing the similarity between photonic and phononic crystals. We predict that cavities created by modulating the geometry of dispersion-engineered waveguides can support a pair of phonon modes with arbitrarily small frequency splitting, analogous to optical cavities~\cite{Mahmoodian:2010-25693:OE}. Monitoring of this splitting can be applied for sensing. On the other hand, the time scale of mode beating, defined as the inverse of frequency splitting, can be tuned for resonant regimes of nano-manipulation.

This paper is organized as follows. In Sec.~\rsect{modes}, we present an approach for dispersion engineering in nano-waveguides, and analyze the properties of slow phonon modes associated with split band edges.
Then in Sec.~\rsect{vortex} we establish a correspondence between the slow-phonon vortices and field singularities.
Finally, in Sec.~\rsect{defect} we discuss the features of pairs of phonon defect cavity modes, supported by dispersion-engineered waveguides with modulated geometrical parameters.
We present conclusions in Sec.~\rsect{conclusions}.

\section{Slow phonon modes and dispersion engineering} \lsect{modes}

\begin{figure}
\includegraphics[width=\columnwidth]{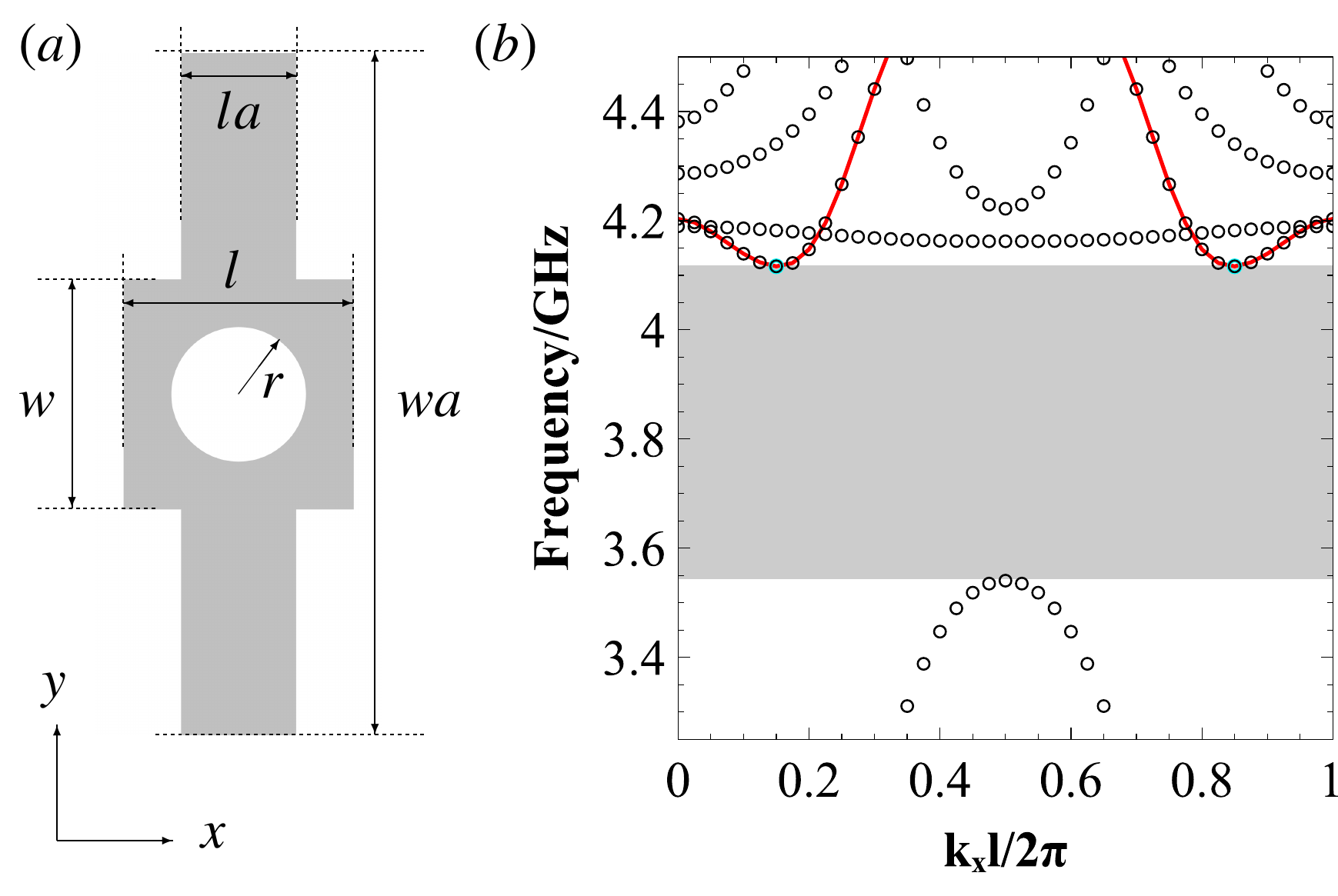}
\caption{\label{fig:structure}
(a)~Schematic top view of the periodic waveguide supercell. 
(b)~Band diagram of eigenmode dispersion: grey shading highlights the band gap, and the split band-edge modes marked by cyan open circles.
The geometry parameters are $l$=$500$nm, $w$=$500$nm, $wa$=$1500$nm, $la$=$250$nm, $r$=$150$nm, and the height of the structure is $h$=$220$nm.
}
\end{figure}

To present the principles of dispersion engineering and slow phonon mode properties in nanowaveguides, we consider
a periodic structure configuration with experimentally accessible geometry following Refs.~\cite{Pennec:2011-41901:AIPA, Gomis-Bresco:2014-4452:NCOM}, see Fig.~\ref{fig:structure}(a) for the supercell. Whereas here we consider the properties of phonon modes, the advantage of this structure is that it can simultaneously provide photonic band-gaps at optical wavelengths.
The structure consists of a silicon strip waveguide with symmetric 'acoustic wings' attached at each side and circular air holes drilled at the centre. The wings provides strong acoustic reflection, such that the periodic waveguide has a very robust band gap.

We investigate the acoustic wave propagation in periodic nano-waveguides using the wave equation for the mechanical displacement $\mathbf{u}$,
\begin{equation}\label{eq:eigenProblem}
    \rho\frac{\partial^2\mathbf{u}}{\partial{t}^2}=
             \bigtriangledown\cdot\left[C:\frac{(\bigtriangledown\mathbf{u})^T
                                          +\bigtriangledown\mathbf{u}}{2}\right] .
\end{equation}
This equation expresses the Newton's second law, that the equilibrium mass density $\rho$ multiplied by the acceleration vector $\partial^2\mathbf{u}/\partial{t}^2$ equals to the elastic restoring force at the right hand side, where $C$ is the stiffness tensor.
We consider lossless regime with no damping.

We seek solutions for periodically oscillating modes with frequency $f$ as $u(x,y,z,t) = u(x,y,z,0) \exp(i 2 \pi f t)$,
\begin{equation}\label{eq:eigenProblemFreq}
    - \Omega^2 \rho \mathbf{u} =
             \bigtriangledown\cdot\left[C:\frac{(\bigtriangledown\mathbf{u})^T
                                          +\bigtriangledown\mathbf{u}}{2}\right] .
\end{equation}
Acoustic eigenmodes in periodic waveguides satisfy the Bloch-Floquet conditions, $\mathbf{u}(x+l,y,z) =\mathbf{u}(x,y,z)\exp(i k_x l)$. In numerical simulations, we apply these conditions at $x=\pm l/2$ cross-sections, and the rest of the boundaries are treated as stress free. We normalize the eigenmode solutions such that the maximum displacement amplitude inside the structure is scaled to unity, $max(|\mathbf{u}|)=1$. The simulations are performed by the eigen-frequency solver in structural mechanics module provided by COMSOL Multiphysics Finite Element Analysis Software. We treat silicon as a cubic material with density $\rho$ = $2330$ $kg/m^3$ and stiffness tensor $C$, which has three independent elastic coefficients $c_{11} = 166$~GPa, $c_{12} = 64$~GPa, and $c_{44} = 80$~GHz~\cite{Pennec:2011-41901:AIPA}.

\begin{figure}
\includegraphics[width=\columnwidth]{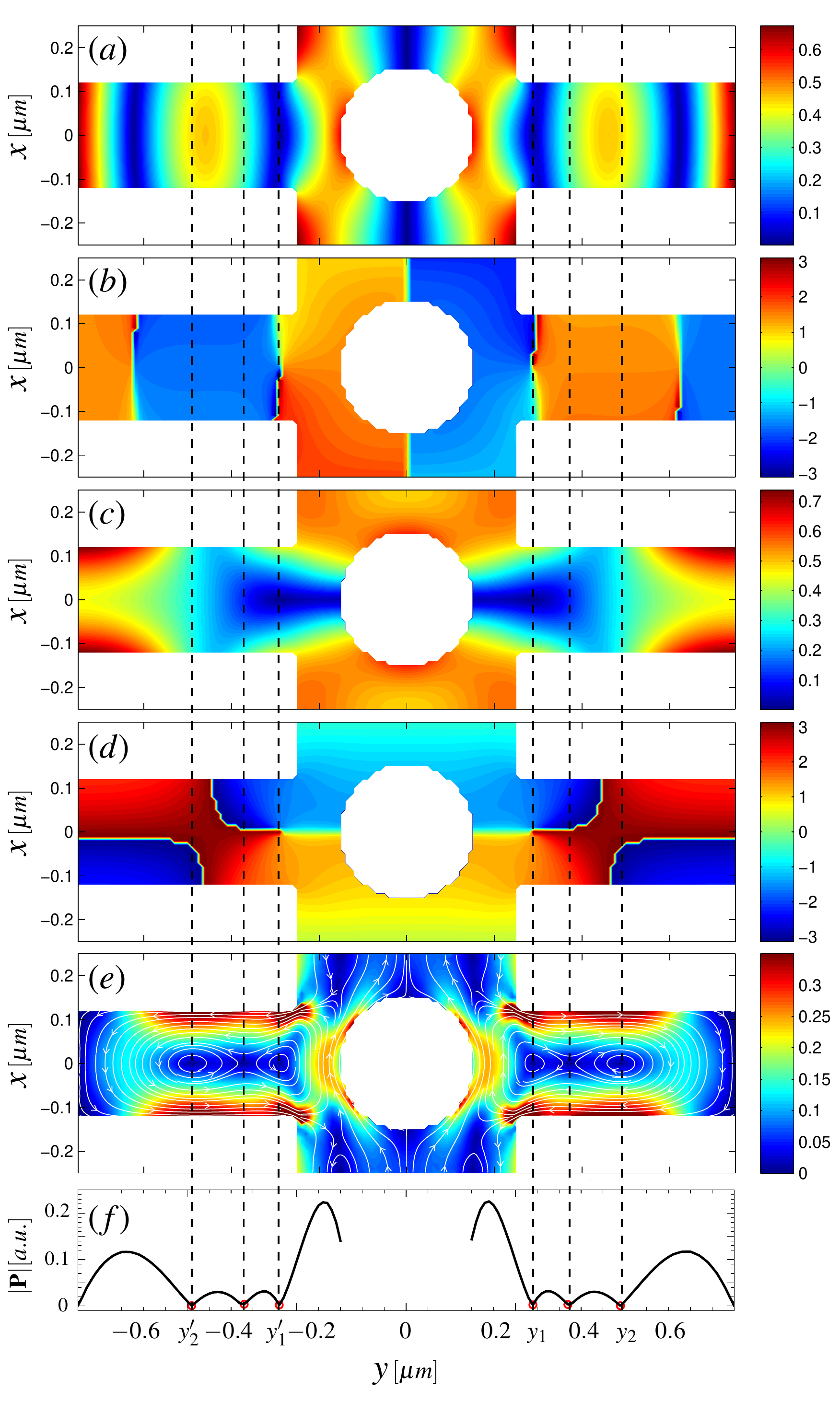}
\caption{\label{fig:modeProfile} 
(a-d)~Mode profile of the split band-edge mode at ${k_xl}/{2\pi}$=$0.15$ in $z=0$ plane. 
(a,c)~Amplitudes and (b,d)~phases of the (a,b)~$u_x$ and (c,d)~$u_y$ displacement components. 
(e)~The time-averaged energy density flux $\langle\mathbf{P}\rangle$ shown with white streamlines on the background color-coded magnitude $|\langle\mathbf{P}\rangle|$ 
(f)~The amplitude of time-averaged energy density flux along $x=0$ line in~(e).
}
\end{figure}

We optimize the structure geometry to obtain the dispersion feature of split band edges (SBE),
%
which are marked by cyan open circles in Fig.~\ref{fig:structure}(b). The corresponding modes have wave-numbers inside the first Brillouin zone, while their group velocities can be vanishingly small close to band-edge. The modes are a mixture of transverse and longitudinal waves as it has its dominant displacement in both $x$ and $y$ directions. We show the normalised $x$ and $y$ components of the displacement mode profile at $k_xl/2\pi=0.15$ in $z=0$ plane in Figs.~\ref{fig:modeProfile}(a-d). The displacement parallel to propagating direction, $u_x$, is symmetric with respect to a plane $x=0$ and anti-symmetric to $y=0$, while the $y$ component is anti-symmetric with respect to $x=0$ and symmetric to $y=0$. It is also worth mentioning that there are clearly phase screw dislocations in both longitudinal ($x$) and transverse ($y$) components, namely at ($x=0$, $y\simeq 0.62\mu$m) in $u_x$ and at ($x=0$, $y\simeq0.29\mu$m) in both $u_x$ and $u_y$. We didn't show the displacement in $z$ direction since it is three orders of magnitude smaller than these two dominant components. For the band-edge mode at $k_xl/2\pi$=0.85 the displacement profile has the same symmetry and magnitude but with an opposite phase.



For all kinds of band-edge modes the group velocity approaches zero, and therefore the total energy flux also has to vanish. However the local energy fluxes can behave differently, as these are linked to the local phase gradients.
A key distinctive feature of SBE modes is the presence of nontrivial phase gradient along the waveguide according to the Bloch-Floquet conditions, in contrast to the band-edge modes where phase can remain flat and only jump by $\pi$.
It has been shown that optical SBE modes in photonic crystals can sustain nonvanishing local energy fluxes at arbitrarily small group velocities~\cite{Sukhorukov:2009-94016:JOA}, and here we extend this analysis to phonon fields.
The energy flow of an acoustic field is defined as
\begin{equation} \leqt{flow}
   \mathbf{P} = -\frac{\partial\mathbf{u}}{\partial{t}}
   \cdot\left[C:\frac{(\bigtriangledown\mathbf{u})^T
                        +\bigtriangledown\mathbf{u}}{2}\right] .
\end{equation}
%
We plot the streamlines of the time-averaged vector on top of its time-averaged amplitude for the SBE mode in Fig.~\ref{fig:modeProfile}(e). We observe that non-zero local energy flow has both forward ($+x$) and backward ($-x$) components at the boundary of $x=\pm{l/2}$, similar to the photonic SBE modes.

Furthermore, the time-averaged energy flux also forms local circular and spiral energy flows. At the centre of such flux vortices, the energy flow density reaches its local minima, see the energy flow across line $x=0$ at $z=0$ plane in Fig.~\ref{fig:modeProfile}(f). In order to clearly illustrate this feature with respect to the phase screw dislocations in displacement fields, we label the local minima with open red circles and position vertical dashed lines at all of the local minima positions in $y$. Positions $y_1(y_1^\prime)$ and $y_2(y_2^\prime)$ give zero amplitude and correspond to the vortex centres, while the middle ones in between represent the saddle points in the flux between two vortices and have positive energy flow. At $y_1\sim0.29\mu{m}$ both $u_x$ and $u_y$ feature phase singularities and they lead to vanishing time-averaged energy flow at the vortex core, whereas at $y_2\sim0.49\mu{m}$ neither $u_x$ nor $u_y$ shows phase screw but the time-averaged energy flow inside vortex is also zero.



We see that the two circulating flows of the same vorticities, e.g. clockwise at $y_1$ and $y_2$, while appearing simultaneously at different spatial positions for the same eigenmode as shown in Fig.~\ref{fig:modeProfile}(e), should have very different structures. Indeed, the first one at $y_1$, as already noted above, corresponds to zero displacement amplitudes and screw dislocations of the same chirality in both linear components of the displacement $u_{x}$ and $u_{y}$. Thus, the energy flux and the kinetic energy, under the restriction of finite energy, naturally vanish at this point. It is not the case for the second type of vortex at $y_2$, as there are no zeros and no phase singularities in the components $u_{x,y}$. Note, however, that there is another phase dislocation in the $u_x$ component at about $y\simeq 0.62\mu$m, visually shifted further to the right from the vortex at $y_2\simeq0.49\mu$m. Although these two features of the displacement profile do not correspond to each other spatially, they suggest an analogy with optical polarization singularities, which enables us to explain the differences in the physical origin of two phonon vortices.

\section{Phonon vortices and field singularities} \lsect{vortex}

We suggest an approach to the analysis of phonon vortices in analogy of similar treatment of transverse optical fields~\cite{Bliokh:2006-174302:PRB}. We recall that the $u_z$ component of the displacement is three orders of magnitude smaller than its in-plane counterparts. These displacements $u_{x,y}$ form a vector field, and we can apply the mathematical approaches developed for the two-dimensional transverse optical fields. Specifically, we can use optical theories by formally treating $u_{x,y}$ mechanical displacement components as the values of linearly polarized components of the field ${\mathbf u}$. The optical energy flow for the latter vector field has been analyzed extensively~\cite{Bekshaev:2007-332:OC, Mokhun:2007-261:OPAP, Berry:2009-94001:JOA, Bekshaev:2011-53001:JOPT} in the context of presence of phase singularities~\cite{Nye:1974-165:PRSA, Soskin:2001-219:PROP} in its components as well as {\em polarization singularities}~\cite{Dennis:2009-293:PROP} in the spatial map of polarization distribution. At each spatial point of this map the state of the field vector $\bf u$ is represented by polarization ellipse, left- or right-handed, depending on the temporal dynamics of $\bf u$. Points in plane where the ellipse is degenerated into a circle, i.e. the points where polarization is purely circular and the orientation of the ellipse is undeterminate (singular), are called C-points. Each C-point corresponds to one of the three topological types of polarization singularities: lemon, star, and monstar~\cite{Dennis:2009-293:PROP}, as well as two types in the contour classification, elliptic or hyperbolic~\cite{Dennis:2002-201:OC}.

\begin{figure}
\includegraphics[width=0.9\columnwidth]{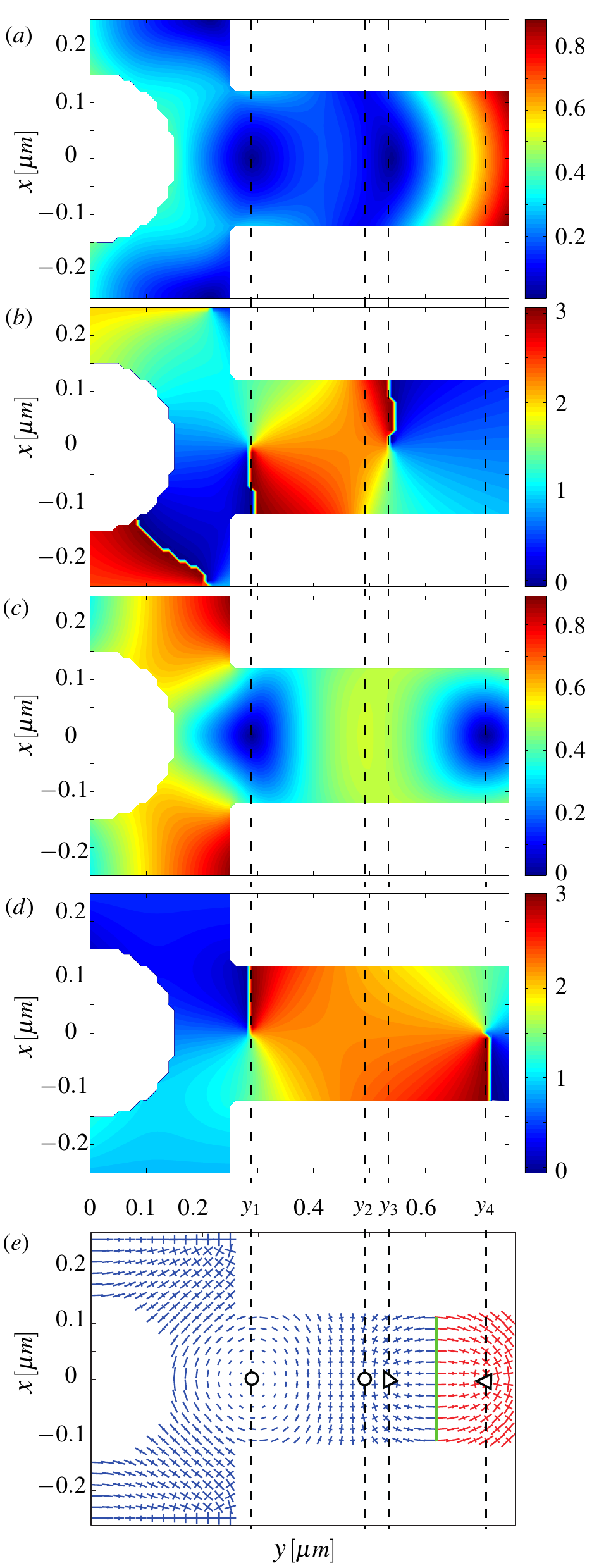}
\caption{\label{fig:map} 
(a-d)~Profiles of the circular polarization components (a,b)~$u_+$ and (c,d)~$u_-$. Show are the (a,c)~amplitudes and (b,d)~phases.
(e)~Polarization map shown with major axes of polarization ellipses; blue and red for left- and right-handed ellipses, respectively. Circles indicate positions of two vortices at $y_{1,2}$ [cf. Fig.~\ref{fig:modeProfile}(e)], and triangles indicate the positions of C-points at $y_{3,4}$.
}
\end{figure}

We introduce the circular polarization components of ${\bf u}=u_+ {\bf c}_+ + u_- {\bf c}_-$, namely $u_\pm = (u_x\pm i u_y)/\sqrt{2}$ and the eigenvectors of circular polarizations are given by ${\bf c}_\pm = ({\bf x}\mp i{\bf y})/\sqrt{2}$, here $\bf x$ and $\bf y$ are Cartesian unit vectors. We plot the amplitude and phase profiles of circular polarization components in Figs.~\ref{fig:map}(a-d). The whole structure is summarized in the polarization map plotted in Fig.~\ref{fig:map}(e), where we also indicate with circles the positions $y_{1,2}$ of two vortices observed in Fig.~\ref{fig:modeProfile}(e).

The first type of vortex at $y_1$ clearly appears as zeros in both $u_\pm$, because the total field vanishes here, ${\bf u}(0,y_1)=0$. The polarization ellipses degenerate into linear segments around vortex at $y_1$, i.e. the polarization is linear at each point in the vicinity of the vortex and it is directed tangential to concentric circles. This is a typical picture for an ``azimuthally polarized'' field~\cite{Oron:2000-3322:APL}.

In addition to this expected defect, we also observe in Figs.~\ref{fig:map}(b,d) phase dislocations at different locations $y_{3,4}$ in two components $u_\pm$, also clearly visible as zeros of corresponding amplitudes in Figs.~\ref{fig:map}(a,c). Polarization map in Fig.~\ref{fig:map}(e) illustrates this complex spatial structure. Let us start with the rightmost zero of the $u_-$ component at $y_4\simeq 0.71\mu$m. Since the $u_-$ component vanishes and $u_+$ dominates, the field is right-handed (red crosses), and the resulting C-point is marked by a left-pointing triangle in Fig.~\ref{fig:map}(e). The topological index $+1/2$ of this C-point can be determined by simply following the direction of the twist by major axis of polarization ellipse when following a counter-clockwise contour around C-point. Calculation of the characteristic determinant~\cite{Dennis:2002-201:OC} reveals the {\em hyperbolic} type of this polarization dislocation.

Another C-point at $y_3$ is left-handed and corresponds to phase dislocation in $u_+$ component. Its topological index is also $+1/2$ but this point is of {\em elliptic} type. The importance of such classification lies in the fact that, in general, while hyperbolic singularities correspond to stagnation points of the current, the elliptic C-points correspond to a vortex energy flow~\cite{Berry:2009-94001:JOA}. Indeed, we find the position of a vortex at $y_2$ is very close to elliptic C-point at $y_3$, and there are no vortical flows around hyperbolic point at $y_4$. Note also that between two C-points there is a line of pure linear polarization, separating ellipses of opposite chiralities, see the green line in Fig.~\ref{fig:map}(e)]. The phase singularity in the $u_x$ component, mentioned above, lies exactly at this so-called L-line; similar structure has been observed in laser beams in birefringent crystals~\cite{Volyar:2006-3724:OE}.

We point out that the vortex near elliptic C-point is spatially separated from it, their positions do not coincide exactly. The reason is that the C-point is noncirculaly deformed~\cite{Mokhun:2007-261:OPAP}. More precisely, such separation is expected when, in addition to the orbital part of the energy current there is also spin current present, the distinction between two contributions into total power flow is most clear in paraxial laser fields~\cite{Bekshaev:2007-332:OC, Bekshaev:2011-53001:JOPT}. The effect of spin current can be opposite to the orbital flow and the vortex is shifted from the position of a deformed C-point, in contrast to the ideal case of a circular C-point at the origin of circular streamlines of a vortex current~\cite{Berry:2009-94001:JOA}. Thus, the splitting of the vortex and C-point positions serves as an indication of the presence of phonon spin currents and may further stimulate the inquiry into interesting effects of Berry phase and spin Hall effects for phonons~\cite{Bliokh:2006-174302:PRB}.

\section{Pairs of phonon cavity modes}\lsect{defect}

\begin{figure}
\includegraphics[width=\columnwidth]{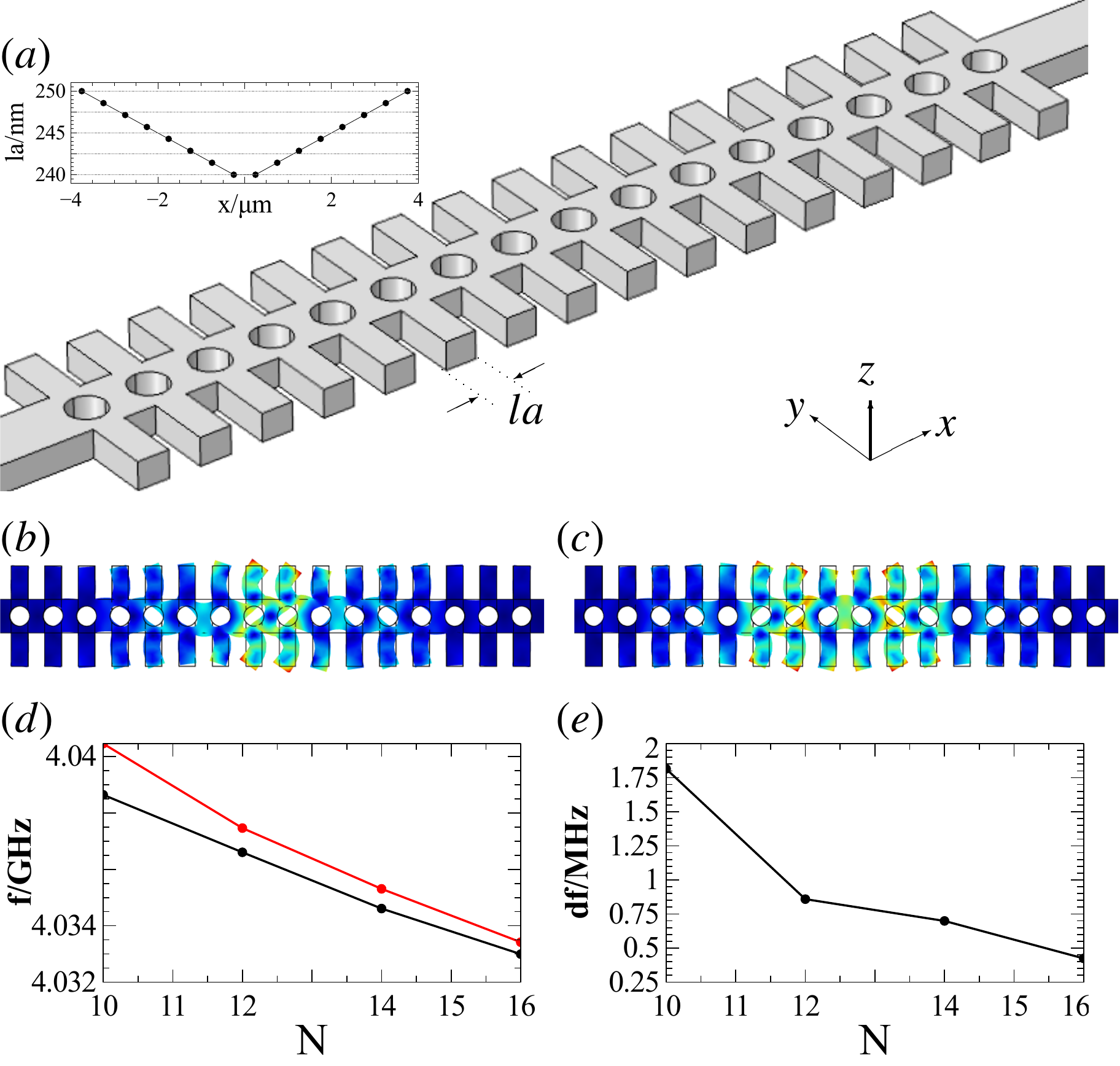}
\caption{\label{fig:cavity} 
Defect phonon modes of a modulated waveguide. 
(a)~3D schematic with linearly tappered wing width $la$, as shown in the inset. 
(b,c)~Top view of the total displacement for defect modes at resonant frequencies (b)~$4.046$~GHz and (c)~$4.048$~GHz. 
(d)~Resonant frequencies and (e)~frequency splitting of two cavity modes vs. the number of tapering holes.
}
\end{figure}

After revealing the similarities and differences between slow phonons and photons of the SBE modes in periodic nanowaveguides, in this section, we examine the phononic defect cavities originating from the SBE modes. In the context of photonic crystal cavities, it was predicted that a pair of defect modes can appear~\cite{Mahmoodian:2010-25693:OE}, and we perform such analysis for phonon fields.

We introduce the defect modes and create the gradual structural modulation by tapering the width of the 'acoustic wings' $la$ linearly from outside ($la=250$~nm) to the center ($la=240$~nm) along $N=8$ periods, see Fig.~\ref{fig:cavity}(a). Such type of structures can be fabricated experimentally with excellent precision~\cite{Gomis-Bresco:2014-4452:NCOM}.
We perform simulations using COMSOL solid mechanics eigen-frequency solver, considering fixed ends and stress-free conditions at the rest of the boundaries.
We find that such structure indeed supports two defect modes which originate from the slow-phonon vortices at the band-edge, see Figs.~\ref{fig:cavity}(b,c) for the total displacement field of mode $1$ and $2$, respectively. We show here only the modes' top view since, similar to their originating SBE modes, the out-of-plane displacements of both the defect modes are substantially weaker than the displacement in other directions. Defect mode $1$ has symmetric $u_x$ and antisymmetric $u_y$ with respect to $x=0$, while defect mode $2$ has antisymmetric $u_x$ and symmetric $u_y$ with respect to $x=0$. These two cavity-mode resonances have their frequencies very close to each other. Defect mode $1$, shown in Fig.~\ref{fig:cavity}(b), oscillates at $4.046$~GHz while defect mode $2$, shown in Fig.~\ref{fig:cavity}(c), oscillates at $4.048$~GHz.

Furthermore, we find that the frequency splitting between the paired defect modes depends on the gradient of defect introduced to the periodic waveguide to create the localized resonance. We show the phononic crystal cavity frequencies and the frequency splitting as a function of the number of periods in the linear taper $N$, see Figs.~\ref{fig:cavity}(d,e). The frequency splitting between the paired defect modes decreases as the defect gradient gets smaller ($N$ gets larger). The frequency splitting of the modes is very sensitive to the defect gradient, analogous to their photonic counterparts~\cite{Mahmoodian:2010-25693:OE}. One can obtain arbitrarily close resonant frequencies by designing the defect gradient involving tapering of several geometrical parameters, such as width of the strip waveguide $w$, length of the 'acoustic wings' $wa$ or all the geometrical parameters together.

\section{Conclusions} \lsect{conclusions}

To summarize, we demonstrated that persistent slow phonon vortices can appear in periodic nanowaveguides, which dispersion is engineered to feature split band edges. We identify two types of vortices, associated with phase singularity and vanising displacement at the vortex core, or with characteristics reminiscent of polarization singularities in optics.
Furthermore, we find that the waveguide modulations lead to the appearance of a pair of cavity defect phonon modes with small and controllable frequency splitting. 
%
These results suggest new opportunities for the development of acoustic and acousto-optic devices, including applications for frequency shifting, sensing, and nano-manipulation.

\begin{acknowledgments}
Authors thank K. Yu. Bliokh for useful discussions. This work was supported by the Australian Research Council (ARC) Discovery Project DP130100086. Numerical simulations were performed with the assistance of resources provided at the NCI National Facility systems at the Australian National University supported by the Australian Government.
\end{acknowledgments}

%
\bibliography{db_SlowPhononVortices_6}

\end{document}